# Colloidal analogues of charged and uncharged polymer chains with tunable stiffness

Hanumantha Rao Vutukuri *, Ahmet Faik Demirörs, Bo Peng, Peter D. J. van Oostrum, Arnout Imhof & Alfons van Blaaderen *
Soft Condensed matter, Debye Institute for Nanomaterials Science, Utrecht University, The Netherlands

Spherical colloids have been used successfully as condensedmatter model systems for studying fundamental aspects of both phase behavior[1] and dynamic processes such as the glass transition.[2] Recently, a growing interest in colloidal particles with more complex shapes and interactions is fueled by applications in self-assembly and advanced functional materials, as well as by the demand for more realistic colloidal model systems for molecular structures.[3] Examples of such colloidal particles include rod-like particles,[4] regular clusters,[5] chiral chains,[6] staggered chains,[7] granular polymers,[8] particles that exhibit inverse dipolar interactions,[9] patchy particles,[10] and step-growth polymerization of inorganic rodlike nanoparticles.[11] The assembly of colloids into polymerlike chains would constitute a significant step forward in the design of colloidal analogues of molecular structures.[3a]

Polymers are typically divided according to stiffness into three different types, namely rigid rods, semiflexible polymers and flexible polymers.[12] Biopolymers (semiflexible) such as microtubules, actin, and DNA have become useful systems for the study of fundamental aspects of polymer physics because they have several advantages over synthetic polymers.[13] However, even for semiflexible biopolymers, current microscopy techniques cannot be used to study the dynamics of the chains on the individual monomer level.[13, 14] Previous work on the preparation of rigid colloidal bead chains was based on using microfluidic trapping of particles in confined geometries in combination with thermal fusing of particles.[15] It has previously been demonstrated that the stiffness of magnetic bead chains can be tuned by controlling the length of specific linker molecules between beads; this type of tuning has been done in the case of linkers based on the streptavidin—biotin interaction.[16] This method relies on the nature of the particular functional group on the surface of the magnetic particles that binds to the linker molecule and involves several additional chemical steps.[16] Additionally, magnetic materials usually have a high density and strongly absorb light, making them less suitable for the study of concentrated systems in real space.

Herein, we describe new methods to produce colloidal particle chains of three stiffness regimes that can be observed on a single-particle level, that is, on the level of the monomers that make up the chain; the chains can even be observed in concentrated systems without using molecular tracers. These methods rely on the following: dipolar interactions induced by external electric fields in combination with long-range charge repulsion to assemble the particles into chains only, and a bonding step to ensure that the particles remain assembled as chains even after the external field is switched off. We can control the length and the flexibility of the chains. Additionally, we demonstrate that our method is generally applicable by using it to prepare several other colloidal polymers, such as block-copolymer chains, which are formed by combining rigid and flexible chains, spherocylinders, which are formed by heating rigid chains, and both atactic and isotactic chains, which are formed from heterodimericparticle monomer units. We demonstrate that the flexibility of the charged chains can be tuned from very rigid (rod-like) to semiflexible (as in the simplified polymer model of beads on a string[17]) by changing the ionic strength. This method can, in principle, be used with any type of colloidal particle. Moreover, our systems can be matched in terms of refractive index and density, so that bulk measurements in real space are possible.

Colloidal particles that have dielectric constant values that are different from that of the solvent, acquire a dipole moment that is parallel to an external electric field. Dispersions of particles with induced dipolar interactions and with long-range screened Coulombic interactions show rich phase behavior.[18]The dispersion structure can be tuned from chains (1D), to sheets (2D), and eventually to equilibrium 3D crystallites by varying the dipolar field strength.[18] However, when the field is turned off, the dipolar interactions no longer exist and so the particles return to a dispersion.

Making use of this behavior in combination with either thermal annealing or seeded growth on inorganic, organic, and hybrid particles, we demonstrate that permanent bead chains can be made starting from different types of colloids.

Suspensions consisting of monodisperse polymethylmethacrylate (PMMA) particles in cyclohexyl bromide (CHB) were introduced into a thin indium tin oxide (ITO) coated electric cell (Figure 1j). Upon application of an external AC electric field, the induced dipole moment in each particle caused the particles to assemble into chains, one particle thick, aligned in the direction of the field in a head-totail arrangement and with a broad distribution in the length of the chains ($E_{rms}$ = 0.20 V/μm, $f$ = 1 MHz where $E_{rms}$ is the root-mean-square electric field strength and $f$ is the frequency). In the low-field regime, the stable structure is a string fluid phase that consists of chains of particles parallel to the field direction and a liquid-like order of particles perpendicular to the field direction.[18] The time scale for chain formation is on the order of a few seconds. In addition, by increasing the applied field strength, the average chain length could be increased until it spanned the entire width of the gap between the two electrodes. The long-range charge repulsions are essential as they also greatly stabilize the individual strings, preventing them from forming sheets by sideways assembly at higher field strengths.[18a] Furthermore, the strings became straighter and stiffer with increasing strength of the dipole-dipole interactions, which suppressed the thermal fluctuations. The entire sample was then heated to 70-75 ºC, which is still well below the glass transition temperature [19] ($T_g$ = 140-145 ºC) of the PMMA, for about 2-3 minutes using a stream of hot air that was much wider than the sample cell. After removing the field, more than 95% of the dispersion was converted to permanent bead chains with a broad distribution in the length of the chains (Fig. 1a). At elevated temperatures the steric PHSA-PMMA comb-graft-stabilizer that is present on the surface of the particles, redistributes such that the particles that are in contact bond (fuse) or became permanently entangled together by the same van-der-Waals forces between the PMMA polymers that keep the uncrosslinked particles together. The chain structure was preserved and stable even after the field was removed; the chains remained well dispersed due to the long ranged repulsion (κσ ≈ 0.8, where σ is the particle diameter (1.4 μm) and κ is the inverse of Debye screening length, a measure for the range of the electrostatic repulsion). Even at higher salt concentrations the dispersion of chains was stable suggesting that the particles were still sterically stabilized. [18b] In Fig. 1a-c, we demonstrate how fine control of the length of the permanent PMMA chains was achieved by varying the distance between the two electrodes; in this fashion more monodisperse chains of the desired length could be produced. We estimated the polydispersity index for a dispersion of rigid bead chains as seen in Fig. 1c to be 1.11. Fig. 1e shows a scanning electron microscope (SEM) image of a single PMMA chain. In a concentrated dispersion of long bead chains, a nematic phase was observed in three-dimensions by means of fluorescently labeled chains with confocal microscopy (Fig. 1d, Movie S1) where the single particles in the chains could be clearly distinguished. Note that a nematic phase possesses long range orientational order, but no long range positional order. Additionally, we could easily obtain the 3D information to calculate, for instance, the orientational order parameter which was determined to be ~0.8 in accordance with a nematic phase.[20] The nematic phase transition was accelerated by an external electric field.

We used our method to make rigid bead chains consisting of different polymeric particles, such as polystyrene (PS)

particles (Figure 1 f), and hybrid particles containing a silica core and a PMMA shell (see the Supporting Information). In addition, the shape of the PS bead chains could be changed by prolonged heating (4 h) at 958C; upon this treatment, the bead chains transformed into spherocylinders (Figure 1 g). A slightly modified procedure was used to produce permanent rigid inorganic colloidal bead chains as well. This modified procedure, which involved the growth of a thin layer of material around the chain, was used to prepare titania-(Figure 1i) and silica-covered spheres (Figure 1 h; see the Supporting Information).

We made further modifications to our simple method to fabricate (semi)flexible chains of dielectric beads. PS bead chains in dimethyl sulfoxide (DMSO) became flexible when the PS beads were sterically stabilized with a higher molecular weight polyvinylpyrrolidone (PVP, $M_w \geq 360$ kg/mol). The longer bead chains exhibited different conformations, which are purely driven by thermal fluctuations as shown in Fig. 2a-d. We quantify the flexibility of the chains by estimating the persistence length ($l_p$) for our rigid (~ 40 mm or 30000 $\sigma$) and semi-flexible (~ 14 μm or 10 $\sigma$, and the Kuhn length ≈ 20 $\sigma$; see SI) particle chains using Fourier mode analysis.[21] The ratio between the persistence length and the contour length ($l_p/l_c$) for our (semi-)flexible strings is on the order of 1, whereas this ratio is larger than 1000 for our rigid rods (Fig. S9). For a comparison, for flexible polymers such as λ-DNA this ratio is much smaller than one (lp ≈ 50 nm, lc ≈ 16 μm). In the case of semiflexible polymers such as actin filaments, the ratio is on the order of one (lp ≈ 16 μm, lc ≈ 20 μm). [21] For rigid polymers such as microtubules the ratio is on the order of 1200 (lp ≈ 6 mm, lc ≈ 50 μm). The distribution of bond angles was calculated for flexible and rigid 8 bead chains as seen in Fig. 2e-f, from which we conclude that the bond angle distribution is Gaussian. By comparing to a Boltzmann distribution it is seen that a bond angle of 20 degrees has an elastic energy cost of *1 $k_BT$*. To further probe the flexibility, a single bead in a chain was trapped with optical tweezers and dragged through the dispersion. Hairpin-like conformations were observed as a result of the viscous drag on the chain (Movie S3). In addition, ring and knot like structures could be made with a single optical trap on a long semi-flexible chain (Fig. 2g-j). Once the optical trap was released, the electrostatic repulsion, the elastic energy and thermal motion caused the chain to relax. We note that when the PS particles were stabilized with the higher molecular weight PVP the chains became flexible (see SI), whereas stiff chains resulted in the case of electrostatically stabilized particles (see SI). Therefore, we believe that the stabilizing polymers (PVP) act as linkers between the beads by being entangled with the PS polymer chains.

More complex chains could be made by applying our method on a dispersion of hetero dimer colloids, which consisted of PMMA ($\sigma$ = 1 μm, rough in the SEM inset Fig. 3a-b) on one side and PS ($\sigma$ = 0.85 μm, smooth in the SEM inset Fig. 3a-b) on the other side that were stabilized with PVP. Strings formed from these particles were also observed to be flexible. Note that the dielectric constants of these two materials are close to each other ($\varepsilon_{pmma}$ = 2.6, $\varepsilon_{ps}$ = 2.4), therefore no preferential dimer ordering was observed within the chain, in analogy to atactic polymers (Fig. 3a, Movie S4). However, controlling the internal structure of a chain would be important in, for example, photonic applications. We therefore exploited the charge asymmetry between both sides of the dimer (PS-PMMA) to obtain control over the internal structure of heterodimer chains. Prior to chain formation, all the particles were first aligned in a particular direction using a low DC field (1.5 V/mm) for about 10-15 secs that caused the side with the most negative zeta-potential to point to the positive electrode (in this case PS side). The particles were then exposed to an AC field of 1 MHz for a much longer time (2-3 mins), followed by a heating step. The resulting heterodimer chains were similar to isotactic polymers. The different refractive indices of both ends of the heterodimeric particles (1.59 for the PS end and 1.49 for PMMA end) allowed the ends to be distinguished using optical microscopy. The darker PS ends are distributed in a regular manner within the chain (Figure 3b); this result is strikingly different from the random arrangement of heterodimeric particles in the previous experiment (lower inset of Figure 3a; see also the Supporting Information, Figure S11). Additionally, an SEM image of the chain clearly shows that the smooth parts (PS) of the chain are regularly arranged (upper inset, Figure 3b); this image can be compared with the SEM image of a chain in the previous experiment, in which the arrangement of smooth parts is random (upper inset, Figure 3a).

An interesting feature of our flexible chains is that the flexibility of already synthesized charged chains can be tuned by manipulating the interactions between the beads in the chain. These interactions were manipulated in three ways: by applying an external electric field (Fig. 2k-l), by varying the range of electrostatic repulsions ($1/\kappa$) (Fig. 2m), and by inducing depletion attractions (Fig. 3a & c). The (semi-)flexible PS bead chains became rigid when the chains were transferred from a normal solvent ($\kappa\sigma \approx 5$, $\sigma = 1.35$ μm) (DMSO) to a strongly deionized solvent with $\kappa\sigma \approx 1$, a value comparable to that for polyelectrolytes [21] (Fig. 2m, Movie S5). The longer ranged electrostatic repulsions between the beads forced them into a straight conformation. Furthermore, we induced short-range attractions by adding a non-adsorbing polymer (1.75 wt% dextran of $M_w$ = 5000 kg/mol), which acted as a depletant to the semi-flexible PS-PMMA dimer chains in water. The attractive depletion interactions between the beads induced a folded state of the semi-flexible chains (Fig. 3c, Movie S6). The conformational states of a polymer are highly dependent on the monomer-solvent and monomer-monomer interactions. The flexibility tuning plays an important role in polymer physics. For instance, Doi has predicted that rotational diffusion of a stiff polymer in a crowded environment is independent of stiffness,[22] whereas Odijk has estimated that such diffusion should be enhanced by flexibility.[23]

We were also able to make chains where a part of a chain was flexible while the remaining part was rigid as in a triblock copolymer (Fig. 3d and Movie S7). We achieved this by mixing flexible and rigid chains together and repeating the same protocol of making bead chains.

In conclusion, we have developed a general method for preparing dielectric bead chains of high purity with a high yield. The method can be applied generally because it relies only on i) a repulsive double-layer interaction that is in the size range of the particles, ii) an induced dipolar attractive force in the direction of an external electric field and iii) short-range attractions that keep the particles together; such attractions include van derWaals forces, linkages formed using molecules such as PVP, and thin layers formed by seeded growth.We also showed that the semiflexible and rigid bead chains behave in a similar way to a variety of polymers and simplified polymer models.[24] Altering the interactions between the colloidal beads, the interactions between bead and solvent, and the molecular weight of the linker (PVP) enables exquisite control over chain conformations and new insights into the behavior of simplified polymer models down to the monomer level in real space. Moreover, we have shown that our method can also be applied to particles of more complex shape and composition. We believe that the method could be applied to smaller particles by using the "giant" electrorheological effect.[25]

[∗] Dr. H.R. Vutukuri, Dr. A.F. Demirörs, B. Peng, Dr. P.D.J. van Oostrum, Dr. A. Imhof, Prof. dr. A. van Blaaderen, Soft Condensed Matter, Debye Institute for Nanomaterials Science, Utrecht University, Utrecht,The Netherlands E-mail: H.R.Vutukuri@uu.nl, A.vanBlaaderen@uu.nl

[∗∗] We thank J.C.P. Stiefelhagen for PMMA & PS particle synthesis. It is a pleasure to acknowledge F. Smallenburg, E. Barry and R. Chelakkot for useful assistance in persistence length calculations. H. R. V is part of the research program of the 'Stichting voor Fundamenteel Onderzoek der Materie (FOM)', which is financially supported by the 'Nederlandse organisatie voor Wetenschappelijke Onderzoek (NWO)'.  H. R. V, A. F. D and B. P were supported by the EU (project Nanodirect).

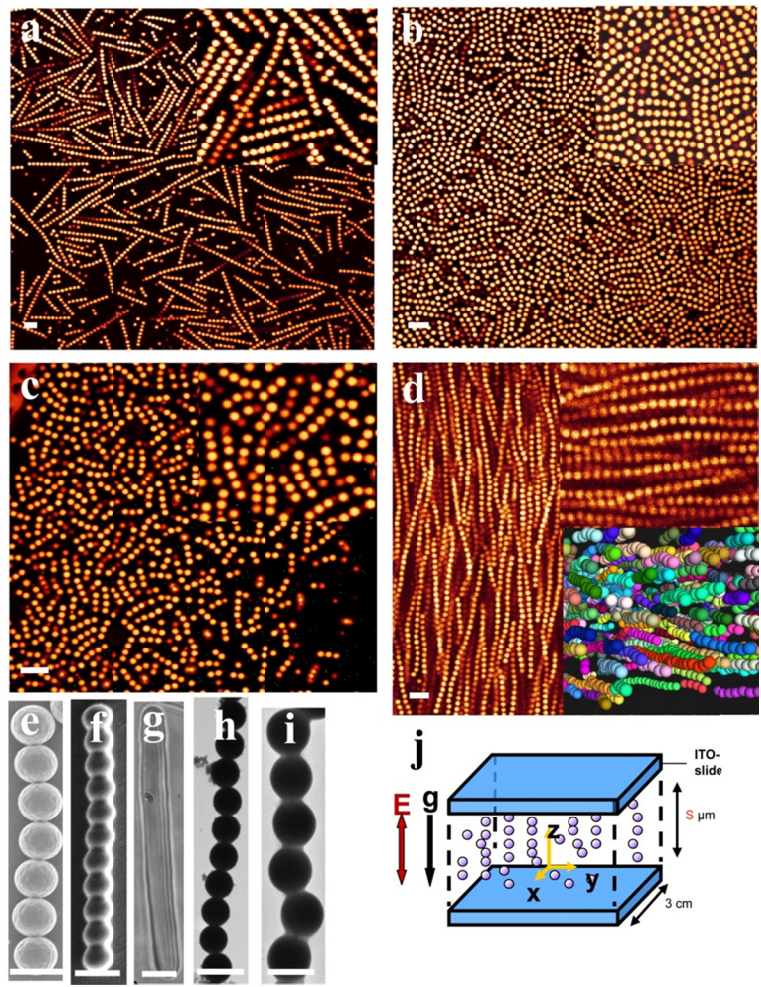

**Figure 1.** Permanent rigid colloidal bead chains. a–c) Confocal micrographs of permanent PMMA chains in CHB where the chain length was controlled by varying the distance ($S$) between the two electrodes; the distances were 100 μm, 10–15 μm, and 5–10 μm, respectively; the upper insets are magnified views of the bead chains. d) Confocal micrograph of the nematic phase of long bead chains of PMMA in CHB; upper inset is a magnified view of the micrograph, which has been rotated 90º; lower inset is a rendered representation of particle coordinates and reveals the three-dimensional structure. e–f) Scanning electron microscopy (SEM) images of permanent bead chains of PMMA and PS, respectively. g) Optical micrograph of a spherocylinder of PS, which was prepared by a prolonged heating of bead chains in DMSO. h–i) Transmission electron microscopy (TEM) images of permanent bead chains of silica (SiO2) and amorphous titania (TiO2), respectively. j) Schematic representation of the electric cell. Scale bars: a–d, 5 μm; e–h, 2 μm; i, 1 μm.

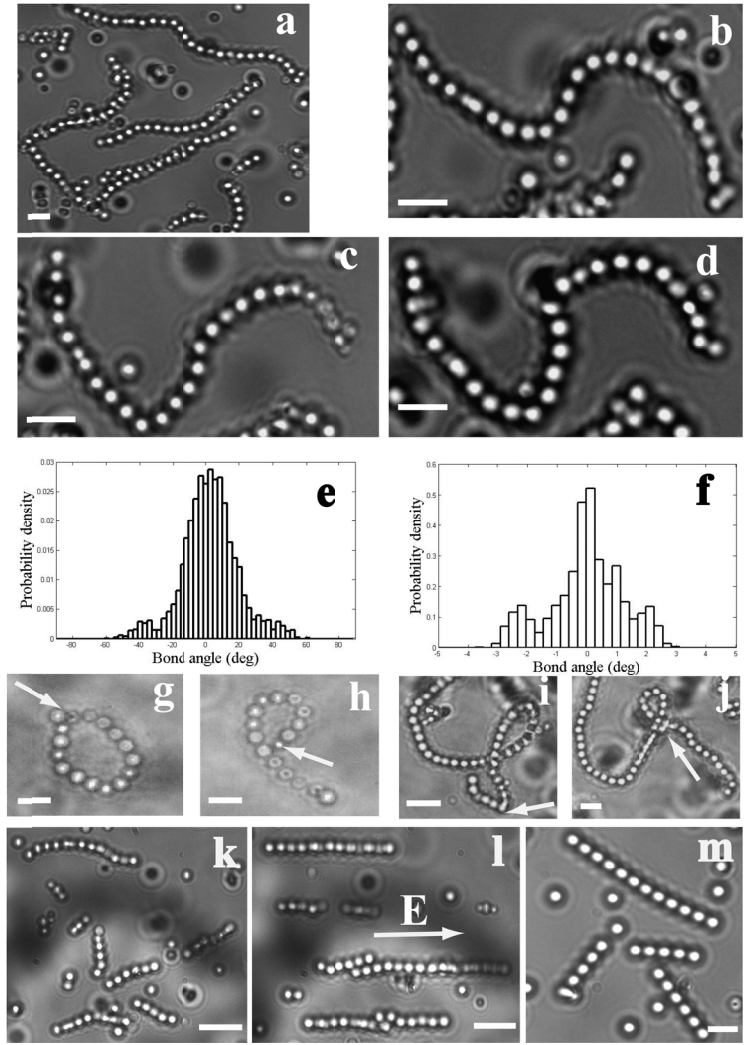

**Figure 2.** (Semi)flexible colloidal bead chains. a) Optical micrograph of PS bead chains in DMSO. b–d) Optical microscopy images of a long chain that is confined to the glass surface by gravity in DMSO taken at different times, a series of images that show the different conformations. e–f) The distribution of bond angles for a flexible and a rigid 8-bead chain, respectively. g–j) Optical micrographs of PS bead chains, in which a single bead in a chain was trapped with optical tweezers (indicated by an arrow) and dragged through the dispersion, thus forming a ring- and a knot-like structure. k–l) Conformations of PS chains in DMSO, in the absence and in the presence of an external field ($E_{rms}$ = 0.06 V/ μm, $f$ =1 MHz), respectively. m) (Semi)flexible PS chains in deionized DMSO ($\kappa\sigma \approx 1$, $\sigma$ = 1.35 μm). Scale bars: 4 μm.

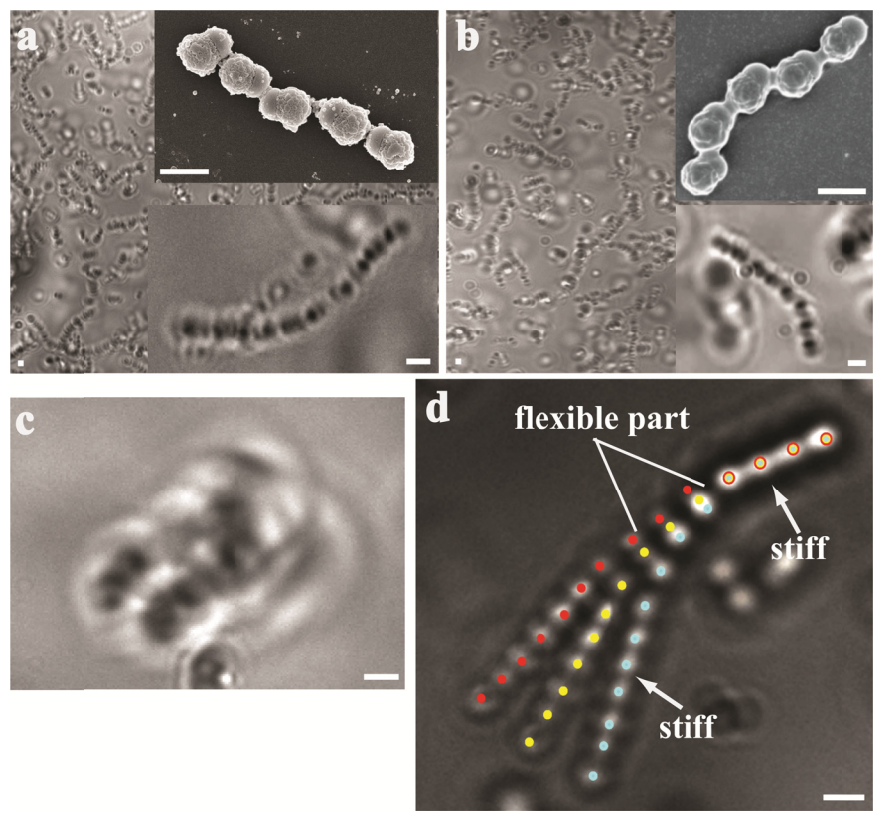

**Figure 3.** Complex chains. a–b) Optical micrographs of a permanent semiflexible atactic-like and isotactic-like chain of PMMA–PS heterodimers, respectively, in water; upper insets are SEM micrographs of the heterodimer chains; lower insets are magnified optical images of the heterodimer chains; the arrows indicate the smooth (PS) parts in the SEM images and the dark parts (PS) in the optical images (see the Supporting Information, Figure S11). c) Optical image of a single chain of heterodimers in the presence of depletion attractions. d) Color-coded overlay of optical micrographs of a triblock-copolymerlike chain of PS beads taken at different times; the overlay was constructed by placing the rigid end in the same position and orientation. Scale bars: 2 µm.